\documentclass[conference]{IEEEtran}
\IEEEoverridecommandlockouts
\usepackage{cite}
\usepackage{amsmath,amssymb,amsfonts}
\usepackage{algorithmic}
\usepackage{graphicx}
\usepackage{textcomp}
\usepackage{amsmath}
\usepackage{array}
\usepackage{multicol}
\usepackage{rotating}
\usepackage{adjustbox}
\usepackage{multirow}
\usepackage{makecell}
\usepackage[table,xcdraw]{xcolor}
\usepackage{tabularx} 
\usepackage{booktabs} 
\usepackage{multirow} 
\usepackage{siunitx} 
\def\BibTeX{{\rm B\kern-.05em{\sc i\kern-.025em b}\kern-.08em
    T\kern-.1667em\lower.7ex\hbox{E}\kern-.125emX}}
\begin{document}

\title{A Portable and Cost-Effective System for Real-Time Air Quality Monitoring and Environmental Impact Assessment}

\author{\IEEEauthorblockN{1\textsuperscript{st} S M Minhazur Rahman}
\IEEEauthorblockA{\textit{Dept.of ECE} \\
\textit{North South University}\\
Dhaka, Bangladesh \\
minhazur.rahman3@northsouth.edu}
\and
\IEEEauthorblockN{2\textsuperscript{nd} Md. Amrin Ibna Hasnath}
\IEEEauthorblockA{\textit{Dept.of ECE} \\
\textit{North South University}\\
Dhaka, Bangladesh \\
amrin.hasnath@northsouth.edu}
\and
\IEEEauthorblockN{3\textsuperscript{rd} Rifatul Islam}
\IEEEauthorblockA{\textit{Dept.of ECE} \\
\textit{North South University}\\
Dhaka, Bangladesh \\
rifat.islam@northsouth.edu}
\and
\IEEEauthorblockN{4\textsuperscript{th} Ahmed Faizul Haque Dhrubo}
\IEEEauthorblockA{\textit{Dept.of ECE} \\
\textit{North South University}\\
Dhaka, Bangladesh \\
ahmed.dhrubo@northsouth.edu}
\and
\IEEEauthorblockN{5\textsuperscript{th} Mohammad Abdul Qayum}
\IEEEauthorblockA{\textit{Dept.of ECE} \\
\textit{North South University}\\
Dhaka, Bangladesh \\
mohammad.qayum@northsouth.edu}
}

\maketitle

\begin{abstract}
Air pollution remains a major global issue that seriously impacts public health, environmental quality, and ultimately human health. To help monitor problem, we have created and constructed a low-cost, real-time, portable air quality monitoring system using cheap sensors. The system measures critical pollutants PM2.5, PM10, and carbon monoxide (CO), and environmental variables such as temperature and humidity. The system computes the Air Quality Index (AQI) and transmits the data via a Bluetooth connection. The data is relayed, in real time, to a mobile application. Because of its small size and low manufacturing cost the system readily lends itself to indoor and outdoor use and in urban and rural environments. In this paper we give an account of the system design, development, and validation, while demonstrating its accuracy and low-cost capabilities. We also consider its wider environmental, social, and regulatory implications with regards to; improving public awareness, being used for sustainability purposes and providing valuable information for informed decision making.
\end{abstract}

\begin{IEEEkeywords}
Air Quality Monitoring (AQI), PM2.5, PM10, Carbon Monoxide, Portable Sensors, Real-Time Monitoring, Environmental Sustainability, Public Health, Wireless Data Transmission
\end{IEEEkeywords}

\section{Introduction}
One of the most significant environmental challenges of our time is air pollution. A great deal of research and innovation has gone into trying to ensure access to clean and breathable air. The first step in ensuring the clean air we breathe is our ability to predict air quality as well as monitor how the Air Pollution Index (API) changes over time. Several significant models for forecasting air quality have been established; there are many options available, including a range of methods, both classical, and deep learning models for air quality forecasting.

Urbanization and its accelerating pace confront people with many challenges in many facets of life. Urbanization challenges have (and will continue to) affect people's lives: transportation challenges, healthcare challenges, air quality challenges (to name a few). Air quality degradation has now become a significant issue in many areas. The World Health Organization (WHO) reported that air pollution is responsible for more than seven million deaths each year; shockingly, around 80\% of the urban population occupy areas in excess of WHO's air quality recommendations\cite{b1}.

Bangladesh, a South Asian country with a high population density, has witnessed impressive economic development and population growth in recent years, now with a population of over 162 million people. However, this growth contrasts sharply with the serious environmental degradation that has occurred. From the analysis of air quality data from previous years, it is evident Bangladesh is facing alarming air pollution levels. Bangladesh was declared the most polluted country in the world, and the capital, Dhaka, was number 21 among the most polluted cities in the world in 2019\cite{b2}.

Publications utilized traditional statistical models to predict air quality. These models like multiscale air quality model, autoregressive moving average (ARMA), autoregressive integrated moving average (ARIMA), or time series regression models can predict air quality. Bangladesh demonstrated a PM2.5 concentration of 83.30 $\mu g/m^3$ in 2019, within the unhealthy air quality range of 55.5 to 150.4 $\mu g/m^3$. PM2.5 is an abbreviation for particulate matter with an aerodynamic diameter of 2.5 micrometers ($\mu$m) or less which can penetrate deep into the respiratory tract and pose serious health risk. PM2.5 is the largest contributor when calculating pollution levels, with other contributors: PM10 and ozone (O$_3$).

In Bangladesh, air pollution comes mainly from the brick making industry. There are approximately 2,000 traditional brick kilns in and around Dhaka and about 5,200 total nationwide\cite{b3}. Many of these kilns burn coal to create and make bricks for the growing construction sector. Due to the obsolete design and inefficient processes, brick kilns are generally toxic to the air due to the sheer quantities of PM2.5 released and other pollutants during coal combustion activities. So, the brick-making sector contributes roughly about 30 percent of the air pollution that impacts air quality, which harms public health.

Exposure to air pollution is an important contributing factor and risk factor in the initiation and exacerbation of health issues, and where does one start? First, air pollution damages the lungs and lung function, and then there is asthma and COPD (chronic obstructive pulmonary disease); then there are infectious respiratory diseases and finally lung cancer. Exposure to the lungs in polluted air has also shown a significantly higher risk of cardiovascular events such as heart attacks, coronary artery disease, and stroke, especially in urban areas with high traffic and developing populations that experience these same health complications\cite{b4}.

\section{Literature Review}
With air pollution becoming more common, many researches are being done in the hope of creating a valid, cheap, quick, and effective way to observe air pollution. This section will examine existing work around air quality monitoring systems. Liu, Wang, and Co. have designed an indoor air quality monitoring system capable of collecting, uploading, and displaying information including temperature, humidity, and PM2.5 concentration\cite{b5}. The system was built using an Arduino microcontroller, a Si7021 sensor for temperature and humidity, and a GP2Y1010AU0F sensor for detecting PM2.5 particles. They used the ZigBee platform for their communication. Vinod and Co.\cite{b6} have designed a real-time air pollution detection and prediction system for smart cities, similar to intelligent monitoring, when pollution is detected. The system consists of wireless sensors to measure pollution levels in PPM and upload it to the cloud. The system consists of an Arduino board, MQ7 sensor to detect carbon monoxide, GP2Y1010AU0F for detecting dust particles, and CO$_2$ sensor. The system also include a display panel with real-time readings for everything being monitored.

Devarakonda, Sevusu, and their colleagues put forth two low-cost, real-time air quality monitoring systems: one for public transport and the other for personal transport\cite{b7}. The public transport system has a custom mobile sensing unit that includes an Arduino, with CO + PM sensors, a GPS receiver, and a modem. The personal sensor device (PSD), called NODE, has a CO sensor. Both systems transmit geo-tagged data to a server. Both systems visualize pollution in real-time through a web interface. JunHo et al.\cite{b8} proposed an IoT-based indoor air quality monitoring platform consisting of a sensing device, "Smart-Air," and a web server. Leveraging IoT and cloud computing technologies, the device can measure aerosol concentration, VOCs, CO, CO$_2$, and temperature-humidity values. This device is very useful for monitoring air quality indoors and was validated by using methodologies from the Ministry of Environment, Korea. They also developed a mobile application to support air quality monitoring. Somansh and Ashish\cite{b9} presented a simple and cost-effective real-time air quality monitoring capability for PM2.5, CO, CO$_2$, temperature, humidity, and air pressure as a standalone unit. They integrated the sensor with IoT, where cloud computing is involved in developing an efficient way to manage sensor data, wherein their low-priced, low-power, and ARM-based Raspberry Pi minicomputer is involved.

Yu-Lin and his group defined the major components of IoT and lithium 2 IC comparison for RFID, M2M, and sensor networks for RFID sensor networks and propose an intelligent multi-functional monitoring platform designed to spatially and temporally reduce pollution in the air\cite{b10}. Yu-Lin and Farmer's comparison identified the effectiveness of integrated systems based on networked communication, cloud decision making, information tracking, and online management in air quality improvements. Mohieddine et al. described an end-to-end Intelligent Air Quality Monitoring (IAQM) system that could measure CO$_2$, CO, SO$_2$, NO$_2$, O$_3$, Cl$_2$, and ambient temperature and relative humidity\cite{b11}. Their system is built around a local gateway to connect a set of wireless sensor nodes to the internet from which access to data was remote. Reliable dissemination of data were emphasized and there are mechanisms for data backup and recovery in response to internet outages. The IAQM architecture adopted the open-source IoT platform Emoncms, did support live air quality monitoring, long term storage, and integrated into numerous diverse sensors and wireless technologies.

Helton and Gustavo\cite{b12} presented an IoT-based system based on inexpensive sensors to monitor pollutants that are harmful to human health, using WHO rules. The device uses PMSA003, MICS-6814, and MQ-131 sensors to identify pollutants such as PM2.5/PM10, ozone, carbon monoxide, nitrogen dioxide, and ammonia. The device is powered by an ESP-WROOM-32 microcontroller that integrates Wi-Fi and Bluetooth in order to transmit data to the cloud. Vanessa et al.\cite{b13} reported portable environmental monitoring devices based on IoT embedded in vehicles. These devices provide data by using MQTT to transmit data to a server with an edge-based time series database. The cloud is engaged in data visualization, with categories taken into account for Low, Normal and High City Air Pollution Levels. A data analysis pipeline of outlier detection and supervised classification was applied to sample air quality in Ibarra, Ecuador. Their system was identified to have more than 90\% inference accuracy with respect to low memory and low power. Swati and others created a three-phase air pollution monitoring system\cite{b14}. The IoT kit uses gas sensors to develop an application via an Arduino IDE and then uses a wi-fi module to reflect the gas sensors; the kit can be placed in various locations in an urban geography. The sensor data via Arduino IDE is uploaded to the cloud. An Android application called IoT-Mobair was created to allow users to access data on air quality. In addition, the application can predict pollution levels while travelling, and warn of dangerous pollution levels when they reach a certain point. 

Moursi and her co-workers presented an IoT-enabled system for monitoring and predicting PM2.5 concentrations at the edge and cloud level\cite{b15}. Using several machine learning algorithms, they developed a hybrid predictive architecture allowing them to use their model as a PM2.5 predictor, especially Nonlinear AutoRegressive with eXogenous input (NARX). They fed the model previous 24-hour PM2.5 levels, wind speed, and rainfall to predict the same hour's PM2.5 level. Their model used evaluation metrics for performance measures including the RMSE, NRMSE, R², IA, and execution time showed that the model can be implemented in a low-bandwidth or remote environment. Using an IoT device measuring temperature, humidity, PM10 and PM2.5, Claudia, Adrian and Razvan developed a comprehensive system for assessing the Air Quality Index (AQI)\cite{b16}. The system transmitted data from 5869 data points across the six important parameters to the ThingSpeak cloud. A TensorFlow regression model was used to predict AQI on a real-time basis. The shared study employed feedforward neural networks using both 'adam' and 'RMSprop' optimizers, as well as random forest regression. As a result, the research concludes that the random forest model with 100 estimators is the best model. The Mean Absolute Error (MAE) for AQI10 was 0.2785 and for AQI2.5 was 0.2483 and much can be done to help improve public health through the ability to potentially warn the public when pollution is at its peak.

\section{Background Study}
The term air pollution denotes the dangerous materials in the air. These are pollutants, some of which are gases, particulates and chemical compounds. The material can be either naturally occurring or man-made. Nowadays, it is vital for humans to determine the quality of the air where they live. It should be evident to them whether it is contaminated or good quality. Air pollution was rated four on the world death risk scale. Adequate quality of air is an issue for world governments and individuals. Air pollution is associated with particulate matter from motor vehicles, equipment, waste reclamation, industrial processes and residential areas. There's heavy metal dust, carbon monoxide (CO), ozone (O$_3$), carbon dioxide (CO$_2$), nitrogen dioxide (NO$_2$), hydrogen fluoride (HF), sulfur dioxide (SO$_2$) and there are suspended particulates are also noteworthy pollutants. The health implications of air pollution are serious and include heart attacks, pneumonia, global warming, climate change and many others. Given air pollution's negative implications to human health, environment and general well-being, it is regarded as a serious environmental issue. Air Quality is a reflection of many air quality indices in many jurisdictions.

The Air Pollutant Index (API) is utilized in Malaysia to assess air quality\cite{b17}. In the United States and China, they utilize the Air Quality Index (AQI); in Canada and Hong Kong, they use the Air Quality Health Index (AQHI); in Singapore, they apply the Pollutant Standards Index (PSI), and in Europe, they have a Common Air Quality Index (CAQI)\cite{b18}. For air quality monitoring, Bangladesh uses the Air Quality Index (AQI). At present, there are eleven (11) continuous weather and air monitoring stations that are measuring air quality according to the national standards of five pollutants - NO$_2$, CO, Ozone (O$_3$), SO$_2$, and Particulate Matters (PM2.5 and PM10) in 8 cities in Bangladesh. Air pollution is one of the significant threats we are facing today. Given that air quality has direct consequences on the environment and humans, air quality monitoring is essential. In order to provide the next generations with safer futures, controlling air pollution is now imperative. We can commendably try to lessen air pollution's effects, but we also need to first understand air pollution patterns. 

We must also be careful and limit what we emit as hazardous particulates to the atmosphere. Developing a lightweight, cheap, real-time air quality monitoring device is in this direction. A small device was built to accomplish this which provides an AQI value and is also able to sense harmful gases such as CO and air particles that are pollutants (PM2.5, PM10) that may be present in our surroundings. It also indicates the temperature and humidity of that area. Subsequently, using a Bluetooth module, the device will be able to provide real-time monitoring of air pollution and display the information on a Bluetooth mobile device or a desktop computer. This device is capable of identifying pollutants that are harmful to the environment and people.

\section{Purpose and Goal of the Project}

The main objective of this project is to build a cost-effective, portable, and easily accessible air quality monitoring system.
Specific objectives:
\begin{itemize}
    \item  To design a reasonably priced air quality monitoring system that various individuals can afford.
    \item  To give individuals access to real-time data on air quality, enabling them to learn about the most recent facts regarding contaminants and environmental circumstances.
    \item  To provide user-friendly interfaces, such as those seen in mobile applications, that make it simple for individuals to obtain and comprehend data on air quality.
    \item  To educate the general public on the significance of air quality and its influence on the environment and human health.
    \item  The system should be easily scalable, allowing for the use of many detectors in various environments, including urban and rural areas.
    \item  The project intends to enhance the environment and public health by offering data-driven decision assistance and early warning systems
    \item  The project aims to save the collected data to a memory card for further data analysis approach.
\end{itemize}

\section{Methodology}
Figure 1, is the visual concept of how our machine is designed to take inputs through various sensors, process them in the Arduino, and give output to multiple mediums to represent it to the users. In this figure 2, we can see the circuit diagram of our machine. It shows how all modules are connected in the Arduino Mega and the internal wiring.

\begin{figure*}[t]
\centering
\includegraphics[width=1\textwidth]{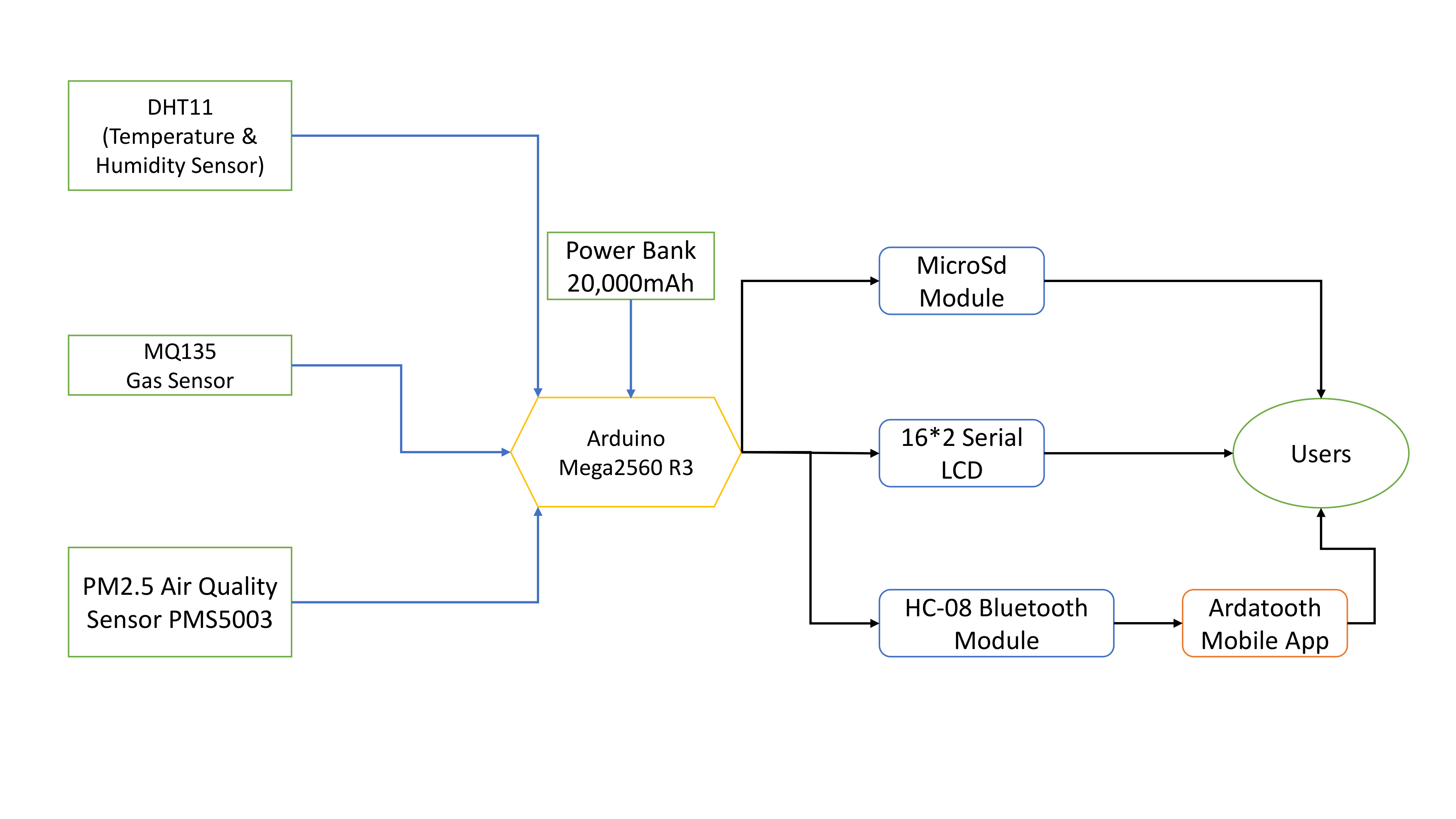}
\caption{Architecture of the proposed prototype design.}
\label{fig:jv}
\end{figure*}

\begin{figure}[htbp]
  \centering
  \includegraphics[width=1\columnwidth]{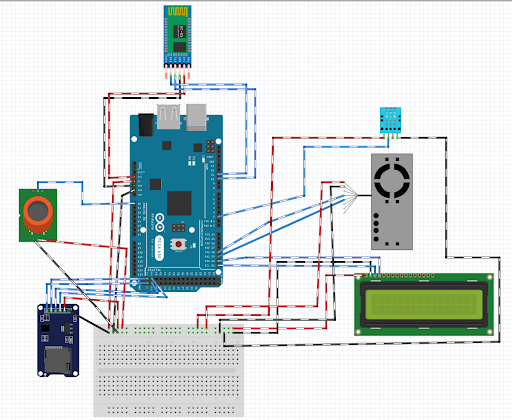} 
  \caption{Circuit Diagram of the device.}
  \label{fig:j}
\end{figure}

\subsection{Hardware and Software Components and Implementations}
We constructed our instrument with Arduino Mega. Air quality sensor Our instrument is comprised of three sensors, of which the PM sensor could detect particles, the MQ135 gas sensor, and the DHT11 for temperature and humidity environmental data. These sensors with Arduino Mega 2560, and real-time data from the sensors is captured till power connection is there, when the sensor produces data with power bank. Also, data can be transfered over Bluetooth connections to an Android phone or via micro USB from the PC. The user has to connect our instrument via Bluetooth interface of Android phone (maintaining) an android application called ArduTooth which sends data to android device). So, the corresponding 16*2 LCD again diplays the values for fast verification.

\begin{figure*}[t]
\centering
\includegraphics[width=1\textwidth]{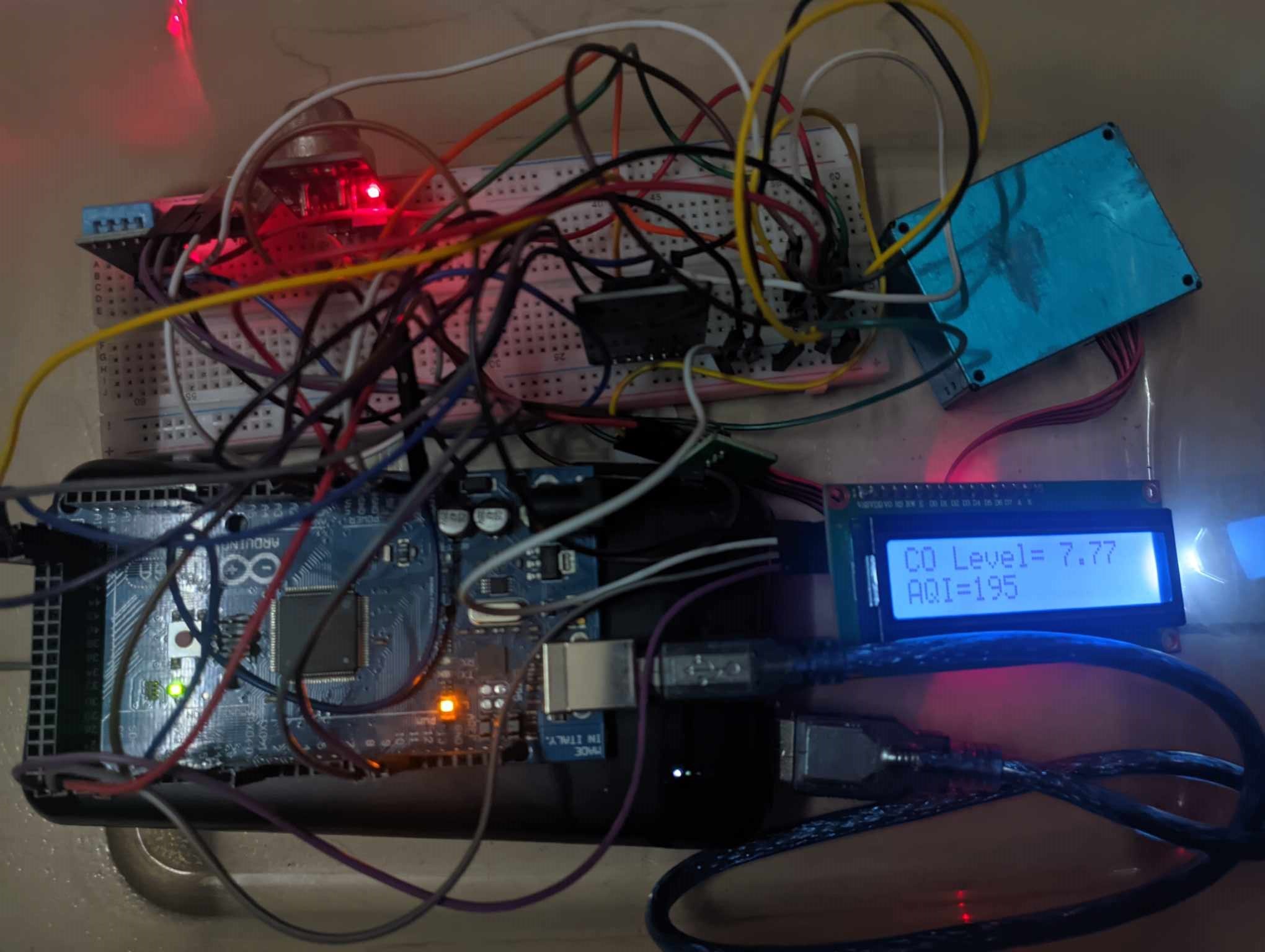}
\caption{Architecture of the proposed prototype design.}
\label{fig:jv}
\end{figure*}

The hardware components we used in our prototype of figure 3 are:

\begin{itemize}
    
\item Arduino Mega 2560: The Arduino Mega 2560 is an excellent microcontroller board with great features and performance. It comes with the entire toolchain to build program and system development including all evaluation kits, C compilers, macro assemblers, in-circuit emulator, and program debuggers or simulators. It has 54 digital input/output pins, 16 analog inputs, 4 UARTs, a 16 MHz crystal oscillator, USB port, power connector, high-speed ICSP header and reset button. We opted to use this board because it is cost-effective, it is fast, and lastly, it has a high number of input pins we can connect a lot of sensors to.

\item PMS5003 sensor: This particulate matter (PM) sensor provides an excellent, very inexpensive opportunity to monitor air pollution. It detects assorted size particulates (PM1, PM2.5, PM10) from a variety of sources including dust, pollen, smoke, metal, organic, etc. A small fan in the sensor pulls air through and sends it past a laser that measures size and number of particles in the air.

\item DHT11 sensor: The DHT11 is a simple, very inexpensive digital temperature and humidity sensor. This sensor measures the ambient air with a thermistor for temperature and a capacitive humidity sensor for humidity and then outputs a reading. It is relatively simple to use, so we utilized the DHT11.

\item MQ135 sensor: It is possible to only use the MQ 135 sensor to sense harmful gases such as benzene, smoke, and vapours in the air. The MQ 135 sensor can sense many harmful gases. It is small and inexpensive so we used the MQ 135 sensor to sense CO gas in the air.

\item MicroSD module: The micro-SD card module is a simple way to communicate data. The pinout is versatile between some microcontrollers; however, it is direct between Arduino. It allows our project to have a large scale storage.

\item 16*2 Serial LCD with I2C: The device has a blue backlit display for maximum clarity on two lines of sixteen characters. The potentiometer located on the back of the hardware is intended to vary the display brightness contrast. The I2C protocol only requires two I/O pins for control (SDA, SCL).

\item HC-05 module: The HC-05 is a Bluetooth module; it offered a serial port that is converted via Bluetooth serial communication module. It enables wireless data transmission and reception using UART. However, it is a single agent device, which means it can only connect via Bluetooth, to most phones and PCs, not to other slave-only devices, such as keyboards or more HC-05 modules. We leveraged this for wireless communication.

\item ArduTooth App: We used an Android Archive Library called ArduTooth, a small lightweight library that makes a stable Bluetooth connection to an Arduino board straightforward. We used ArduTooth to display our air quality measurements on an Android platform.
\end{itemize}

\section{Result and Analysis}
Table I illustrates the stored data in the SD card via the MicroSD card module, which we can view in Notepad. It indicates that six parameters are saved in the module using the Arduino Mega micro-controller, and they can be analyzed using further analysis approach and machine learning. 
\begin{table}[ht]
\centering
\caption{Overall Model Performance Results}
\label{tab:1}
\begin{tabularx}{\columnwidth}{|*{11}{X|}} 
\hline
No & pm2.5 & pm10 & temperature & humidity & CO &  AQI  \\ \hline
1 & 0 & 0 & 29 & 62 & 5.68 & 65 \\ \hline
2 & 180 & 108 & 29 & 62 & 5.27
& 193 \\ \hline
3 & 209 & 118 & 29 & 62 & 5.03
& 223 \\ \hline
4 & 222 & 127 & 29 & 62 & 4.43
& 237 \\ \hline
5 & 224 & 129 & 29 & 61 & 4.03
& 239 \\ \hline
6 & 220 & 125 & 29 & 61 & 3.83
& 234 \\ \hline
7 & 217 & 127 & 29 & 62 & 3.59
& 231 \\ \hline
8 & 222 & 125 & 29 & 61 & 3.30
& 237 \\ \hline
9 & 225 & 128 & 29 & 61 & 4.03
& 240 \\ \hline
10 & 230 & 129 & 29 & 61 & 3.90
& 245 \\ \hline
\end{tabularx}
\end{table}

The Arduino Mega coding is done using the Arduino IDE software. External libraries are incorporated to implement various sensors. The code is programmed in phases e.g. variable definition, sensor data collection, AQI calculation, output in serial monitor, Bluetooth data sending, output in LCD display, and writing it in SD card. All we need to define all necessary variables and libraries at the beginning; for the data reading from the sensor, the MQ sensor gives analog data, which we have to calibrate to get the Carbon Monoxide (CO) level in the air. PMS5003 sensor gives values of particle matter in air, which is particles less than or equal to 2.5 micrometers (PM 2.5) and 10 micrometers (PM 10) or less. The DHT11 sensor gives values for temperature and humidity in the Arduino, in digital pins.

One of the most important parts of this project is to determine the Air Quality Index (AQI) value. There are five levels of AQI value which indicate air quality of the environment. To calculate the value, we used the following equation:\\

I = ($I_{high}$ - $I_{low}$) / ($C_{high}$ - $C_{low}$) * (C - $C_{low}$) + $I_{low}$ \\

I: the (Air Quality) index \\
C: the pollutant concentration \\
$C_{low}$: the concentration breakpoint that is $\geq$ C \\
$C_{high}$: the concentration breakpoint that is $\leq$ C \\
$I_{low}$: the index breakpoint corresponding to $C_{low}$ \\ 
$I_{high}$: the index breakpoint corresponding to $C_{high}$ \\

\begin{figure}[htbp]
  \centering
  \includegraphics[width=0.8\columnwidth]{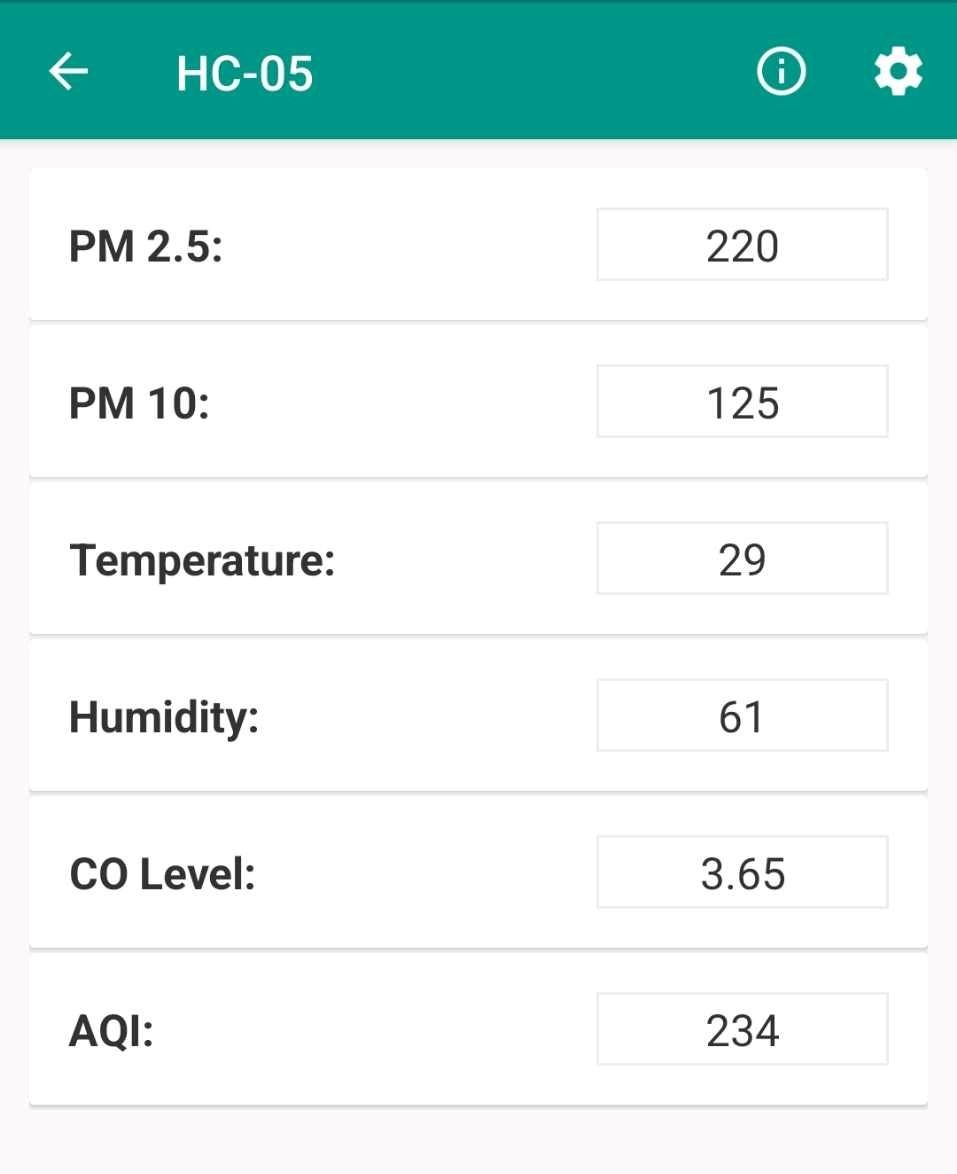} 
  \caption{Ardutooth mobile app.}
  \label{fig:j}
\end{figure}

Figure 4 is a diagram of the Ardatooth mobile application. Data is sent from our machine to the app through the Bluetooth module, which displays the following format. This allows us to see the data in real-time on our mobile phones without any wire connections. 

\begin{figure}[htbp]
  \centering
  \begin{minipage}[b]{0.45\linewidth}
    \centering
    \rotatebox{0}{\includegraphics[width=\linewidth]{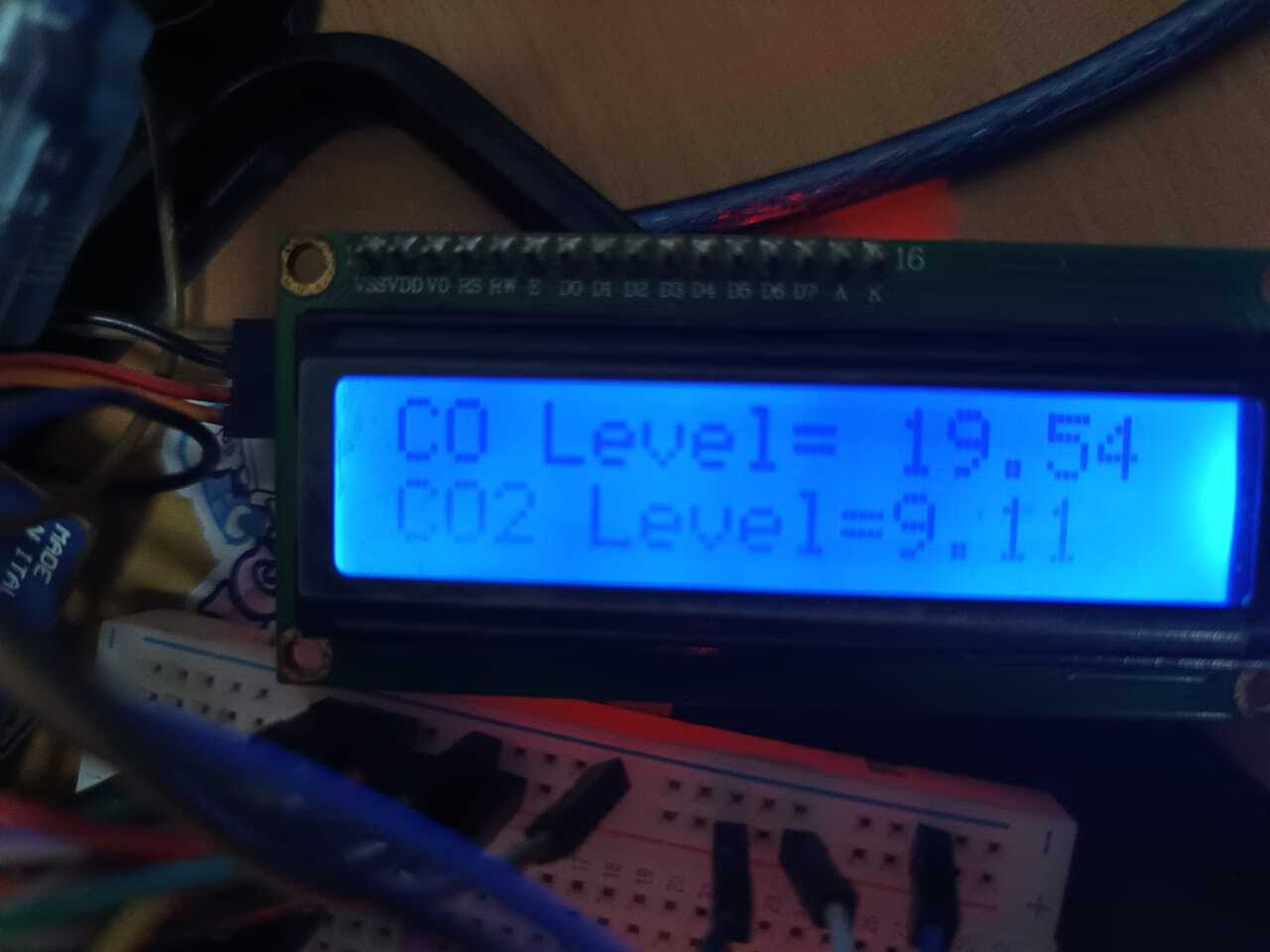}}
  \end{minipage}
  \hfill
  \begin{minipage}[b]{0.45\linewidth}
    \centering
    \rotatebox{0}{\includegraphics[width=\linewidth]{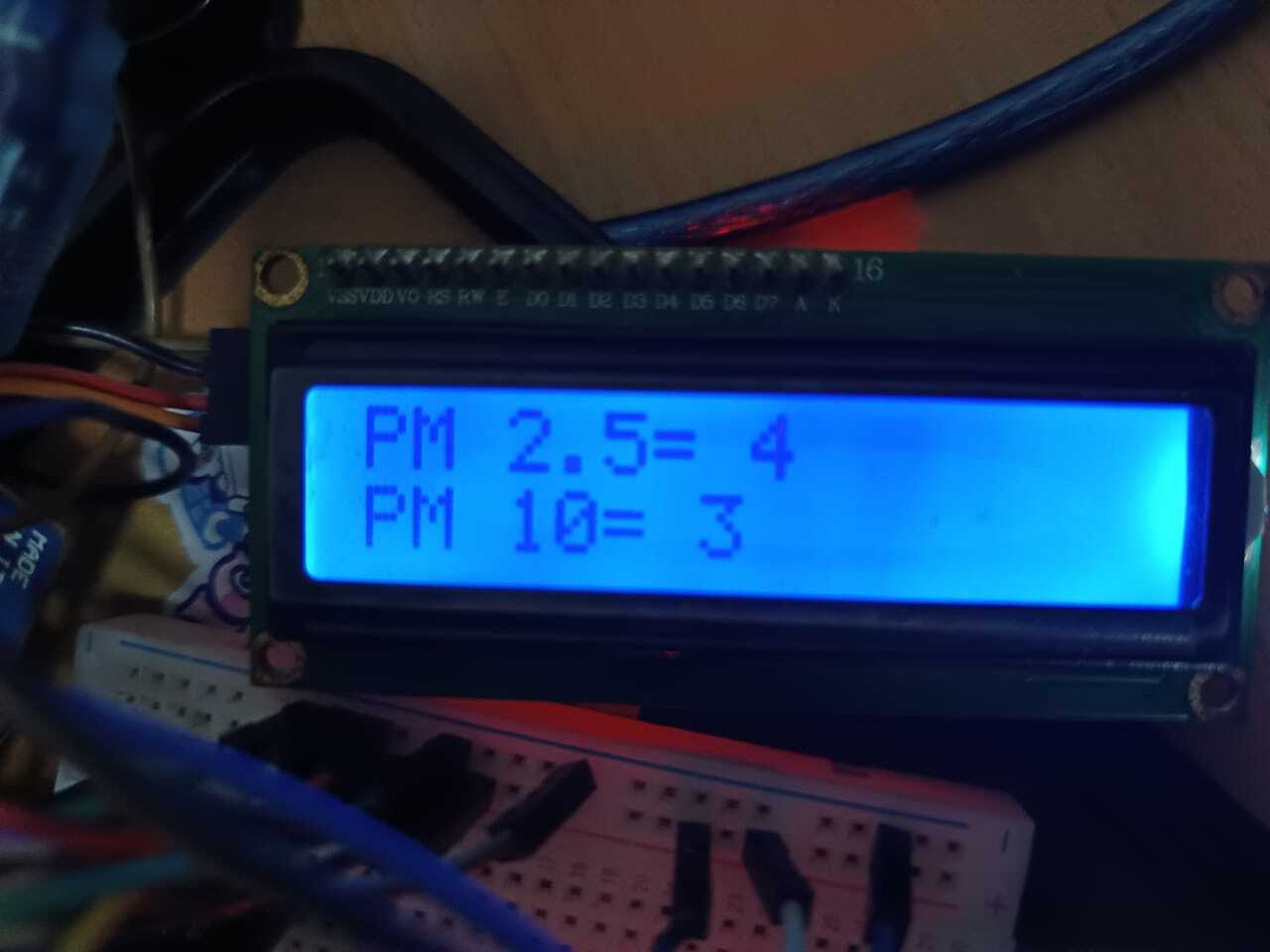}}
  \end{minipage}
  
  \vspace{0.5cm}
  
  \begin{minipage}[b]{0.45\linewidth}
    \centering
    \rotatebox{0}{\includegraphics[width=\linewidth]{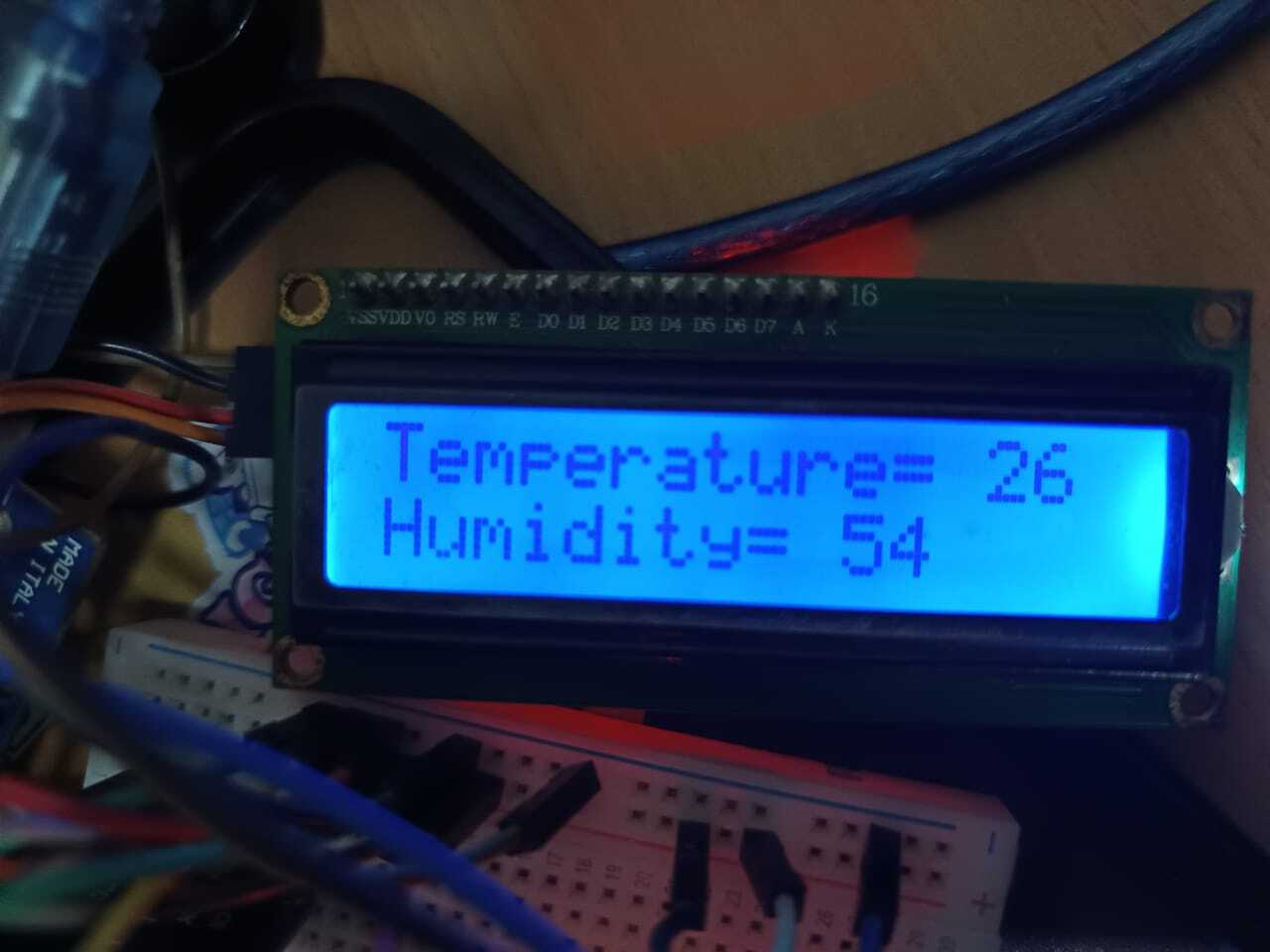}}
  \end{minipage}
  \hfill
  \begin{minipage}[b]{0.45\linewidth}
    \centering
    \rotatebox{0}{\includegraphics[width=\linewidth]{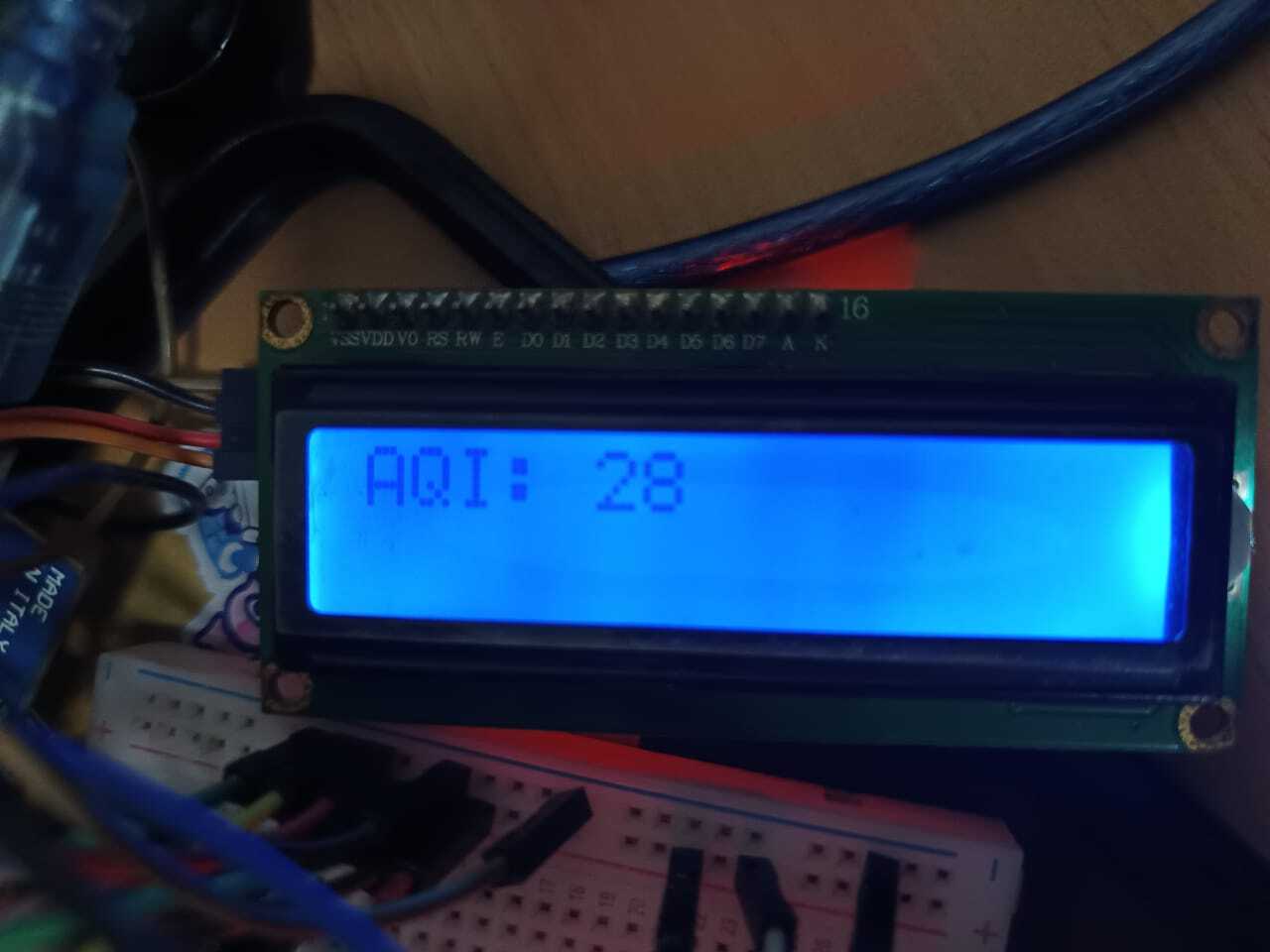}}
  \end{minipage}
  \caption{Outcomes generated by our AQI prototype.}
  \label{fig:comparison}
\end{figure}

To determine pollutant concentration (C), we will be using the CO level, PM 2.5 and PM 10. Upon calculating the AQIs for all the pollutants, we take the max AQI value among that for the final AQI value. Thereafter values are shown on three different platforms like mobile, display, SD card. Figure 5 represents all the outcome results from the prototype machine.

\section{Impact of the Environment}
The Air Quality Index (AQI) is an important tool for monitoring and reporting the current state of air quality conditions. It is one of the most visible and effective instruments used in environmental and sustainability programs by providing the data and now the impetus for actions and regulations to protect our environment and health. The AQI's function is primarily to inform the public about the public's air quality and the associated risks to health and safety. It maps and informs the public about their possible exposures to hazardous air pollutants, which informs people to protect themselves from those exposures. The AQI also helps promote community health and wellness by mitigating those exposures through awareness of and health risks from poor air quality. High levels of air quality indicators often go hand in hand with high levels of pollution and may have serious long-term damages to the environment. The AQI'er in assessing air quality provides information to distinguish the source of pollution, identify the key contributions, and understand the spatio-temporal patterns of pollution in order to frame targeted programs to manage and reduce any emissions or pollutants affecting the areas we are protecting. Clean air is a fundamental but often under-appreciated aspect necessary to ensure biodiversity and ecological health. Biodiversity is inherently tied to clean air and biodiversity represents the cumulative number of species in an environment, which includes not only terrestrial flora and fauna, but aquatic species as well. Air quality is particularly challenging for big cities with urban density and high levels of traffic. 

The AQI can serve as an important guide for urban planners in establishing sustainable cities and transportation systems. Recognizing whether or not higher levels of pollution exist, allows the authorities to invest in needed alterations to the public transport, cycling lanes and pedestrian infrastructure. All of these may minimize pollution levels and increase sustainability. Additionally, the AQI could influence the shift away from fossil fuels toward clean and renewable energy resources. High pollution levels are often from the burning of fossil fuels, and this indicates the need for cleaner alternatives. Through a better understanding of pollution in these speak areas, the policies and incentives in sustainable energy and energy efficiency, assumes a part of cleaner practices initiated from pollution issues identified by the AQI.

\section{Limitation and Future Work}
As with all other equipment, our air quality detection equipment has its limitations. The equipment can be used for indoor and outdoor purposes, and it is portable, but the equipment cannot be used during the rainy season because water will enter very quickly and corrupt the mechanism. We are able to measure air quality from any location, but the equipment cannot measure location, so we must manually enter it. We have gone through numerous modifications and more changes may happen. Database updates for real-time are possible for this work. We have a lot of flexibility for data collection. The data collection can be applied to a machine learning paradigm to determine utilization with decision-making after a lot of time.

\section{Conclusion}
The Air Quality Index (AQI) project is a valuable project in assessing and monitoring air quality in our region. The project has successfully introduced real-time data about multiple air pollutants, enabling a better understanding of environmental conditions following rigorous data gathering, analysis and interpretation. In conjunction with improving community understanding of the importance of air quality, the introduction of the AQI has created opportunities for developing better-informed environmental and public health policy decisions. The watershed success of the project will be measured by its ability to engage communities, develop more sustainable practices, and to improve health outcomes for people. Achieving that, over time, will require ongoing support and expansion of the AQI project, yielding healthier outcomes and a far greater impact in supporting Ontario's air quality standards.

\vspace{12pt}
\color{red}

\end{document}